\documentclass[]{aa} 
 
\usepackage{natbib}

\usepackage{graphicx}
\usepackage[varg]{txfonts}
\usepackage{hyperref}
\def\pc  {(private communication)}
\def\Hdo {$\rm H_{2}O$}
\def\Heu  {\ion{He}{i}}
\def\Cu  {\ion{C}{i}}
\def\Ou  {\ion{O}{i}}
\def\Cu  {\ion{C}{i}}
\def\Pu  {\ion{P}{i}}

\def\Mgu  {\ion{Mg}{i}}
\def\Mgd  {\ion{Mg}{ii}}

\def\CHdou {$\rm ^{12}CH$}
\def\CHtre {$\rm ^{13}CH$}
\def\Teff  {$T_\mathrm{eff}$}
\def\logg  {$\log g$}
\def\loggf {$\log gf$}
\def\vt    {$\rm v_{t}$}
\def\kms   {$\rm km\,s^{-1}$}
\def\hd    {HD\,84937~}
\def\hdd   {HD\,84937}
\def\bd    {BD+44$^{\circ}493$}

\newcommand{\cobold}{{CO$^5$BOLD}}

\begin{document}

\title{
Abundances of the light elements from UV (HST) and red (ESO) spectra in the very old star \hd
\thanks{Based on observations made with the NASA/ESA {\sl Hubble Space Telescope} obtained under program GO-14161 at the Space Telescope Science Institute (STScI), which is operated by the Association of Universities for Research in Astronomy (AURA) and  on observations collected at the European Organisation for Astronomical Research in the Southern Hemisphere (Archives of programmes 080.D-0347(A), 082.B-0610(A), 266.D-5655(A), and 073.D-0024(A) ). }
}
\author {
M. Spite\inst{1}\and 
R. Peterson\inst{2}\and
A. J. Gallagher\inst{1}\and
B. Barbuy\inst{3}\and
F. Spite\inst{1}
 }

\institute {
GEPI, Observatoire de Paris, PSL Research University, CNRS, Universit\'e Paris Diderot, Sorbonne Paris Cit\'e, Place Jules Janssen, 92195, Meudon, France
\and
SETI Institute, 189 Bernardo Ave, Suite 100, Mountain View, CA 94043, USA
\and
Universidade de S\~ao Paulo, IAG, Rua do Mat\~ao 1226, Cidade Universit\'aria, 05508-900, S\~ao Paulo, Brazil
}


\authorrunning{XX1 et al.}

\titlerunning{Abundances of C, O, Mg, Si, P, S, K, and Ca in \hdd}

  \abstract
{ In order to provide a better basis for the study of mechanisms of nucleosynthesis of the light elements beyond hydrogen and helium in the oldest stars, the abundances of C, O, Mg, Si, P, S, K, and Ca have been derived from UV-HST and visible-ESO high resolution spectra in the old, very metal-poor star HD 84937, at a metallicity that is  1/200 that of the Sun's. For this halo main-sequence turnoff star, the abundance determination of  P and S are the first published determinations. 
The LTE profiles of the lines were fitted to the observed spectra. Wherever possible, we corrected the derived abundances for non-LTE effects.  Three-dimensional (3D) \cobold ~model atmospheres have been used to determine the abundances of C and O from molecular CH and OH bands.
The abundances of these light elements in HD 84937 are found to agree well with the abundances in classical metal-poor stars. Our HD 84937 carbon abundance determination points toward a solar (or mildly enhanced) value of [C/Fe]. The modest overabundance of the alpha elements of even atomic number Z, typical of halo turnoff stars, is confirmed in this example. The odd-Z element P is found to be somewhat deficient in HD 84937, at [P/Fe]=--0.32, which is again consistent with the handful of existing determinations for turnoff stars of such low metallicity.
We show that the abundance of oxygen, deduced from the OH band from 3D computations, is not compatible with the abundance deduced from the red oxygen triplet. This incompatibility is explained by the existence of a chromosphere heating the shallow layers of the atmosphere where the OH band, in 3D computations, is mainly formed.
The abundance ratios are compared to the predictions of models of galactic nucleosynthesis and evolution
}


\maketitle

%
\section{Introduction}

The oldest surviving unevolved stars in the Milky Way are made from matter that is not very enriched in elements produced by previous (i.e. older, more massive) stars.
The elements in these surviving stars show low abundances, but the observed deficiencies of individual elements are not identical. The measurement of these deviations is essential in order to constrain the different nucleosynthesis processes producing these elements.  It is therefore very important to measure the detailed abundances for as many elements as possible, with the highest precision, in these very old stars.
This paper is mainly devoted to the abundance of the light elements such as P, S, and K, the evolution of which in the Galaxy is still poorly known.

\hd is a bright, \citep[$V=8.32$,][]{Ducati2002} very metal-poor, main-sequence turnoff star, often used as a reference for the field metal-poor stars \citep[lately:][]{LawlerSC15}, and consequently has recently been widely studied  from spectra in the visible domain 
\citep{GehrenSZ06,LaiJB07,MashonkinaGS11,BergemannLC12,LindMA13,VandenBergBN14,SnedenCK16,AmarsiLA16,ZhaoMY16}. 
The model parameters are very similar in all these studies. The adopted effective temperature \Teff~ ranges from 6300 to 6408\,K, and the metallicity [Fe/H]\footnote{We adopt the classical notation that for element X: $\rm A(X) = log(N_{X} / N_{H}) + 12$ ~and~ $\rm [X/H]=  log(N_{X} / N_{H})_{star} - log(N_{X} / N_{H})_{Sun}$.}
from --1.9 \citep{LaiJB07} to --2.3 \citep{SnedenCK16}.
The surface gravity \logg~ ranges from 4.0 to 4.1 in these studies; the only exception is  a value of  \logg~= 4.5 derived by   \citet{LaiJB07}  from photometry rather than spectra.

Following \citet{VandenBergBN14}, a comparison of the position of the star in the $M_{v}$ vs. log \Teff~ diagram with isochrones computed with different ages and metallicities yields an age of $\rm 12.08 \pm 0.14\,Gyr$, adopting \Teff~= 6408 K and \logg ~= 4.05. But this age depends critically on the adopted temperature. The error bar takes into account only the parallax uncertainty. If we instead adopt the parameters \Teff~= 6300K and \logg~=4.0, as in \citet{SnedenCK16}, then the age of \hd becomes $\rm \approx 13.8~ Gyr$, very close to the age of the Universe: 13.799$\pm$0.038 Gyr \citep{Planck15}. Unfortunately, a new parallax for this star is not yet available in the first Gaia release 
 \footnote{\href{http://www.cosmos.esa.int/web/gaia/dr1}{ http://www.cosmos.esa.int/web/gaia/dr1}}. 

Owing to the weakness of spectral lines at the temperature and metallicity of \hdd, the analysis of very high quality spectra is necessary to establish accurate abundances for this reference star. Taking advantage of a new UV high resolution spectrum from the {\sl Hubble Space Telescope} (HST), complemented by previous high resolution, high signal-to-noise UVES and HARPS spectra obtained in the visible at the European Southern Observatory (Very Large and 3.6m telescopes), we present new determinations of the abundances of several elements with atomic numbers Z lying between Z=6 (carbon) and Z=20 (calcium). New abundances of C, Si, P, and S in particular have been determined from the new HST spectrum, which contains many more useful lines of these elements than the optical and near-infrared spectra do in such weak-lined stars.

Phosphorus (Z=15) has never been measured in \hd and indeed has been measured in very few metal-poor stars, including only five stars with a metallicity below $\rm [Fe/H] < -2.0$\,dex \citep{CaffauBF11,CaffauAK16,RoedererJT14}. Its behaviour 
hints at a primary origin in massive stars \citep{JacobsonTF14}.

Sulphur (Z=16) has never been measured in \hdd, but it has been measured in a sample of metal-poor stars \citep{CaffauBF05,SpiteCA11,CaffauAK16} from near-infrared lines (multiplets 1, 6, and 8) and in a carbon-rich metal-poor star \citep{RoedererPB16} from three UV lines (multiplet 2\,uv)\footnote{With the multiplet numbering of \citet{Moore45}}.
We present the first determination of the sulphur abundance in \hdd, using the UV multiplet 1\,uv compared to the infrared multiplet 1.

For the first time the silicon abundance has been determined in \hd from the profile  of six UV lines and from the Si line in the visible.   

\section {Observational data} 
The HST spectrum was obtained as a legacy spectrum under GO-14161, with R. Peterson 
as PI. As described further by \citet{PetersonKA16}, this spectrum was acquired with the E230H grating of the STIS echelle spectrograph 
and the 0.2" x 0.09" slit for a resolving power R=114\,000. Adopting five of 
the six E230H ``prime'' wavelength settings, with central wavelengths 201.3\,nm, 226.3\,nm, 
251.3\,nm, 276.2\,nm, and 301.3\,nm, resulted in an analysis coverage of 188.0\,nm -- 314.0\,nm. 
The total exposure time of 106.4ks  (29.5 hours) yielded S/N >= 50 per two-pixel 
resolution element at all wavelengths, except for the bluest and reddest.

To compare the abundances deduced from the HST spectra and the visible, we also used ground-based spectra from the ESO Very Large Telescope archives (spectrographs UVES and HARPS). For wavelengths shorter than 400\,nm we used HARPS spectra. They have a resolving power R=115\,000 and a signal-to-noise per pixel (S/N) better than 150 per pixel.
At longer wavelengths, in the blue we combined the individual UVES spectra to a mean UVES spectrum with a resolving power R=65\,000 and a S/N per pixel better than 500.
In the red, in order to minimise the influence of the telluric lines, and taking advantage 
of the fact that following \citet{CarneyL87} and \citet{debruijneE12}, \hd shows no detectable velocity variability, we used three different mean spectra each obtained by combining individual spectra with almost the same geocentric radial velocity. Therefore, the positions of the telluric lines relative to the stellar lines are almost the same. 
These three mean spectra 
were obtained consecutively on a single night every year between 2002 and 2004.
The resulting mean red spectra reach a resolving power of about 45000 and a S/N per pixel equal to 310, 510, and 350, respectively,  in the region of the red \Cu\ lines.

\section {Spectral analysis and abundance measurements} \label{analys}
In our analysis we used OSMARCS model atmospheres \citep {GustafssonEE08} together   with  the {\sf turbospectrum} LTE synthesis code \citep{AlvarezP98,Plez12}. In this code, the collisional broadening by neutral hydrogen is generally computed following the theory developed by \citet{AnsteeO91}, \citet{BarklemPO2000} and \citet{BarklemO2000}.

For the computations we adopted the model parameters and [Fe/H] derived for \hd in the same 
way as in \citet{PetersonK15} and \citet{PetersonKA16}: \Teff=6300K, \logg=4.0, and \vt=1.3\,\kms, along with the iron abundance they derived with these parameters, $\rm[Fe/H]=-2.25$. These values deduced only from the UV spectra are in excellent agreement with the values derived for this star by \citet{LawlerGW13} and \citet{SnedenCK16} based on $V-K$ photometry and accurate {\sl HIPPARCOS} parallaxes.

The  parameters and deduced abundance for each atomic line considered for this analysis are listed in Table \ref{abundlines}.
The mean abundances of the different elements are given in Table \ref{abund}.
For the computations of [X/H] we  adopted the solar abundances A(C)=8.5,\footnote{with the classical notation:  $\rm A(X)= log_{10}(N_{X}/N_{H})+12$} 
 A(Mg)=7.54, A(Si)=7.52, A(P)=5.46, A(S)=7.16, A(K)=5.11, and A(Ca)=6.33, 
following \citet{CaffauLS11} or \citet{LoddersPG09}.

We also provide 3D corrections for a selected number of atomic and molecular lines. They were computed from 3D synthetic spectra based on a \cobold\ \citep{FreytagSL12} model atmosphere from the Cosmological Impact of the First STars \citep[CIFIST;][]{LudwigCS09} grid, using the latest version of Linfor3D
\footnote{\href{http://www.aip.de/Members/msteffen/linfor3d}{http://www.aip.de/Members/msteffen/linfor3d}} 
\citep{GallagherSC16}. A grid of 1D synthetic spectra -- based on 1D LHD models \citep{CaffauL07} computed with the same micro-physics as the 3D models -- were fit to the 3D synthetic profiles to determine the 3D correction \citep [see][] {GallagherCB16}.

The 3D and equivalent 1D LHD models selected from the CIFIST grid have stellar parameters $T_{\rm eff}/{\rm log}\,g/{\rm [Fe/H]}={6206}\,{\rm K}/4.0/{-2.0}$, which is currently the model in the grid with stellar parameters closest to the adopted OSMARCS model parameters.

\begin{table}
\begin{center}    
\caption[]{
Line-by-line 1D LTE abundances A(X) from atomic lines. 
The adopted non-LTE corrections to be applied are given in the last column. For S, the non-LTE correction includes the 3D correction.
Air wavelengths are given for wavelengths > 200\,nm and vacuum  
values below.}
\label{abundlines}
\begin{tabular}{rccccrccc}
\hline
Z &$\lambda$(nm)& Ex. Pot. &  \loggf & $\rm A(X)_{LTE}$ & NLTEcorr&  \\ 
\hline
6 &  {\bf C} (\Cu)&      &          &        &  \\ 
 &   193.0905 &   1.26    & -0.13   &  6.30     &$<0.01$\\
 &   199.2012 &   1.26    & -5.74   &  6.47 &$<0.01$\\  
 &   199.3620 &   1.26    & -3.72   &  6.60 &0.01 \\
 &   247.8561 &   2.68    & -0.96   &  6.28 &0.01 \\  
 &   906.2470 &   7.48    &  -0.455 &  6.41 &-0.07\\
 &   907.8280 &   7.48    &  -0.581 &  6.43 &-0.08\\
 &   908.8509 &   7.49    &  -0.430 &  6.43 &-0.08\\
 &   911.1799 &   7.49    &  -0.297 &  6.49 &-0.10\\
8 &  {\bf O}  &               &         &         \\ 
 &   777.1944 &   9.15    &   0.369 &  7.28 &-0.12\\
 &   777.4166 &   9.15    &   0.223 &  7.28 &-0.12\\
 &   777.5388 &   9.15    &   0.002 &  7.28 &-0.12\\
12 &  {\bf Mg} &         &         &         &  \\
 &   279.5528 &   0.00    &   0.100 &  5.55 &0.00 \\
 &   280.2705 &   0.00    &  -0.210 &  5.55 &0.00 \\
 &   457.1096 &   0.00    &  -5.623 &  5.61 & --- \\
 &   470.2991 &   4.35    &  -0.440 &  5.60 &0.04 \\
 &   517.2684 &   2.71    &  -0.380 &  5.59 &0.06 \\
 &   518.3604 &   2.72    &  -0.158 &  5.54 &0.02 \\
 &   552.8405 &   4.34    &  -0.341 &  5.49 &0.01 \\
14 &  {\bf Si} &         &         &         &  \\ 
 &   197.7598 &   0.00    &  -1.309 &  5.66     &$<0.01$\\
 &   198.0618 &   0.01    &  -1.437 &  5.69     &$<0.01$\\
 &   198.3233 &   0.01    &  -1.192 &  5.65     &$<0.01$\\
 &   208.2021 &   0.78    &  -1.900 &  5.63     &-0.01  \\
 &   208.4462 &   0.78    &  -1.880 &  5.62     &-0.01  \\
 &   212.2990 &   0.78    &  -1.840 &  5.57 & 0.00  \\
 &   212.4122 &   0.78    &   0.149 &  5.40 &$<0.01$\\
 &   390.5523 &   1.91    &  -1.041 &  5.70 & 0.05  \\
15 &  {\bf P}  &         &          \\ 
 &   213.5469 &   1.41    &  -1.240 &  2.98 &  --- \\  
 &   213.6182 &   1.41    &  -0.111 &  2.94 &  --- \\
 &   215.2939 &   1.41    &  -0.357 &  2.81 &  --- \\
 &   253.3986 &   2.32    &  -1.114 &  2.86 &  --- \\ 
 &   255.4911 &   2.33    &  -1.231 &  2.85 &  --- \\ 
16 &  {\bf S}  &         &          \\
 &   190.0287 &   0.00    &  -3.709 &  5.30 &  --- \\
 &   191.4697 &   0.05    &  -4.261 &  5.31 &  --- \\
 &   216.8884 &   1.14    &  -3.962 &  5.30 &  --- \\ 
 &   921.2863 &   6.53    &   0.420 &  5.33 & -0.3 \\
 &   922.8093 &   6.53    &   0.260 &  5.49 & -0.3 \\
 &   923.7538 &   6.53    &   0.040 &  5.40 & -0.3 \\
19 &  {\bf K}& &           &        \\
 &   766.4899 &   0.00    &   0.149 &  3.46 & -0.2 \\
 &   769.8964 &   0.00    &  -0.154 &  3.40 & -0.2 \\
20 &  {\bf Ca} &         &         &         &  \\
 &   428.3011 &   1.89    &  -0.220 &  4.55     & 0.07 \\
 &   431.8652 &   1.90    &  -0.210 &  4.49     & 0.06 \\
 &   442.5437 &   1.88    &  -0.360 &  4.50     & 0.07 \\
 &   443.5679 &   1.89    &  -0.520 &  4.52     & 0.07 \\
 &   445.4779 &   1.90    &       0.260 &  4.50 & 0.04 \\
 &   526.5556 &   2.52    &  -0.260 &  4.63     & 0.13 \\
 &   534.9465 &   2.71    &  -0.310 &  4.50     & 0.12 \\
 &   558.1965 &   2.52    &  -0.710 &  4.67     & 0.09 \\
 &   558.8749 &   2.52    &       0.210 &  4.62 & 0.08 \\
 &   559.0114 &   2.52    &  -0.710 &  4.67     & 0.09 \\
 &   560.1277 &   2.52    &  -0.690 &  4.69     & 0.09 \\
 &   585.7451 &   2.93    &       0.230 &  4.50 & 0.08 \\
 &   610.2723 &   1.88    &  -0.790 &  4.51     & 0.05 \\
 &   612.2217 &   1.89    &  -0.320 &  4.52     & 0.00 \\
 &   616.2173 &   1.90    &  -0.090 &  4.54     & 0.00 \\
 &   643.9075 &   2.52    &   0.470 &  4.42 &-0.02 \\
\hline    
\end{tabular}
\end{center}
\end{table}

\begin{table}
\begin{center}    
\caption[]{Mean abundances of the different elements. 
In Cols. 2 and 3 are given the abundances of the different elements derived from the 1D computations. 
We adopted the solar abundances given in section \ref{analys} and [Fe/H] = --2.25.
In the last column is given [X/Fe] corrected for NLTE and 3D when possible.}
\label{abund}
\begin{tabular}{lccccccc}
\hline
Element        & [X/H]  & [X/Fe]&$\sigma$& [X/Fe]    \\ 
               &  1D    &   1D  &        & corrected \\       
               &  LTE   &  LTE  &        & (adopted) \\
\hline
C (CH)         &--1.86  &  +0.39&  --    &   +0.11   \\
C (\Cu)(UV)    &--2.09  &  +0.16& 0.15   &   +0.17   \\
C (\Cu)(red)   &--2.06  &  +0.19& 0.04   &   +0.10   \\             
&  \\
O (OH)         &--1.47  &  +0.78&        &           \\ 
O (\Ou)(red)   &--1.48  &  +0.77&  --    &   +0.65   \\
&  \\
Mg             &--1.99  &  +0.26&  0.07  &   +0.30   \\
&  \\
Si             &--1.90  &  +0.35&  0.06  &   +0.38   \\
&  \\ 
P              &--2.57  & --0.32&  0.07  &  --0.32   \\
&  \\
S (UV)         &--1.83  &  +0.42&  0.03  &   +0.42   \\
S (red)        &--1.75  &  +0.50&  0.08  &   +0.20   \\
&  \\
K (red)        &--1.68  &  +0.57&  0.03  &   +0.37   \\
&  \\
Ca             &--1.78  &  +0.47&  0.08  &   +0.53   \\
\hline    
\end{tabular}
\end{center} 
\end{table}

\begin{figure}[h]
\resizebox{\hsize}{!}                   
{\includegraphics {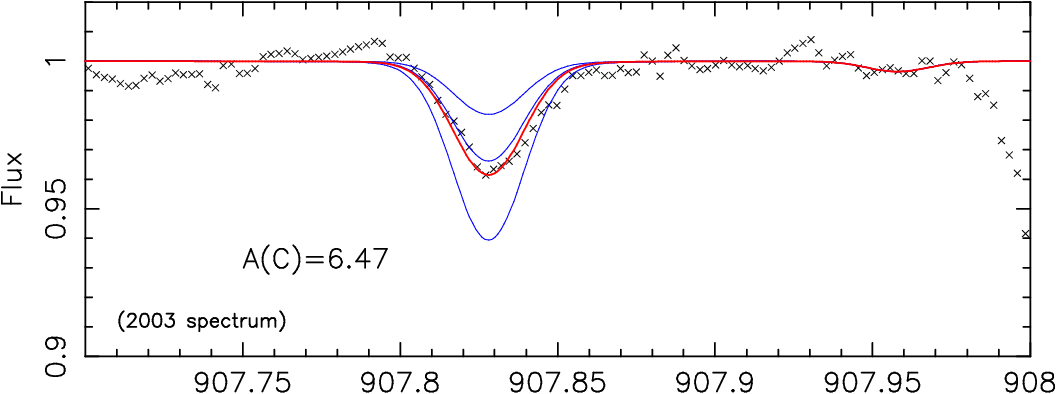}}
\resizebox{\hsize}{!}                   
{\includegraphics {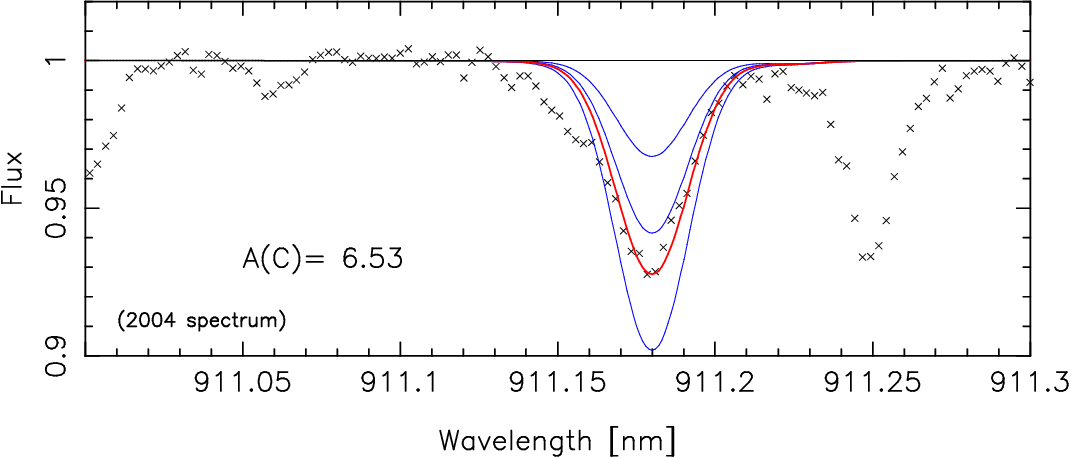}}
\caption[]{Examples of the best fits (red thick line) of two atomic lines of \Cu\ in the red. Synthetic profiles (blue thin lines) have been computed with A(C)= 6.1, 6.4, and 6.7.
}
\label {C1red}
\end{figure}

\begin{figure}[h]
\resizebox{\hsize}{!}                   
{\includegraphics {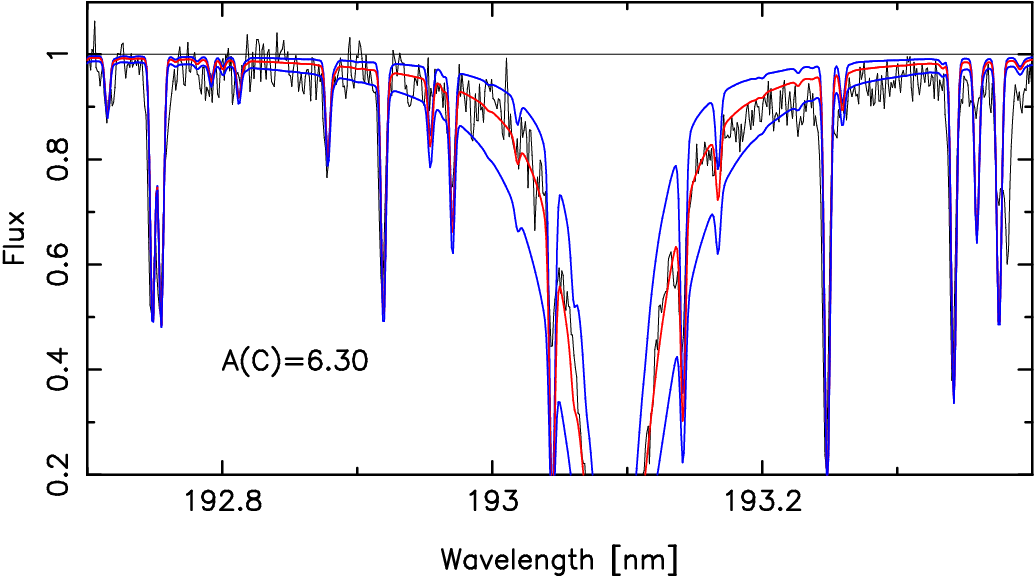}}
\caption[]{Observed profile of the strong \Cu\ line in \hdd. The blue thin lines represent the synthetic profiles computed with A(C)= 6.0 and 6.9. The thick red line represents the best fit (1D computations) A(C)=6.3.}
\label {C1strongUV}
\end{figure}

\begin{figure}[h]
\resizebox{\hsize}{!}                   
{\includegraphics {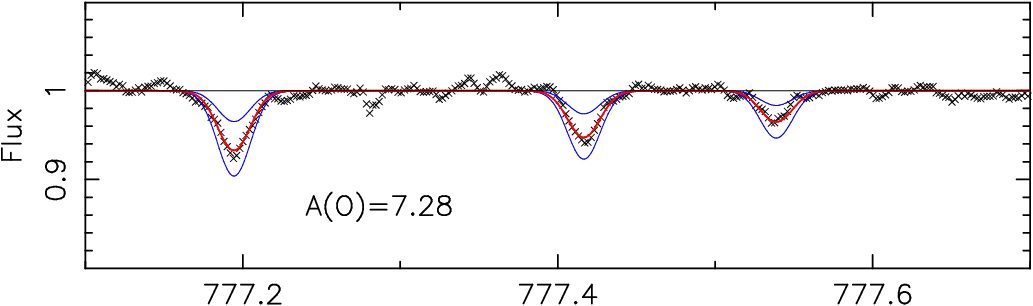}}
\caption[]{Comparison between the observed and synthetic profiles of the red oxygen triplet in \hdd. The synthetic profiles  (blue thin lines) of the oxygen triplet have been computed with A(O)= 6.9 and 7.5. The thick red line represents the best fit for A(O)=7.28. }
\label {Olines}
\end{figure}

\begin{figure}[h]
\resizebox{\hsize}{!}                   
{\includegraphics {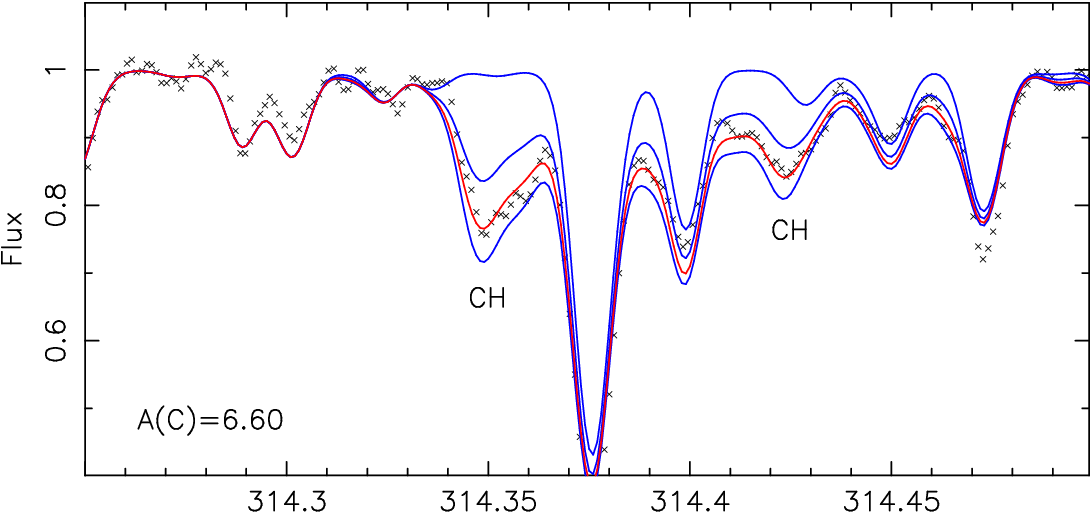}}
\resizebox{\hsize}{!}                   
{\includegraphics {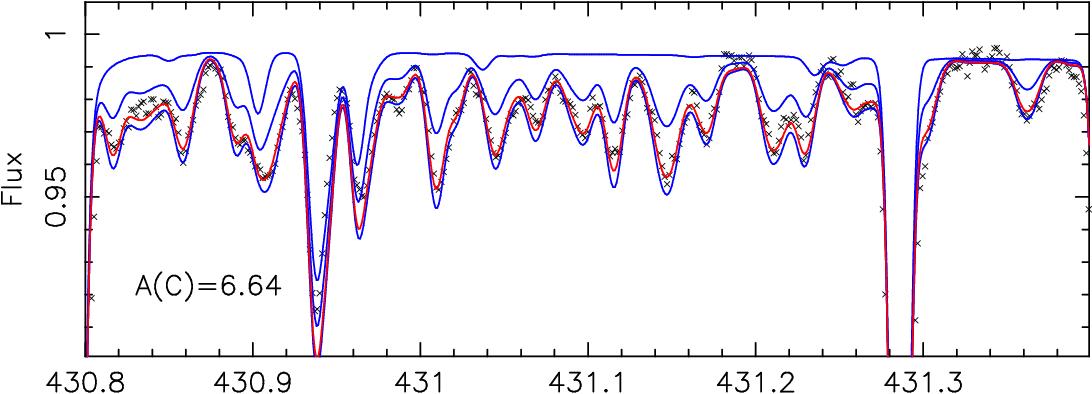}}
\resizebox{\hsize}{!}                   
{\includegraphics {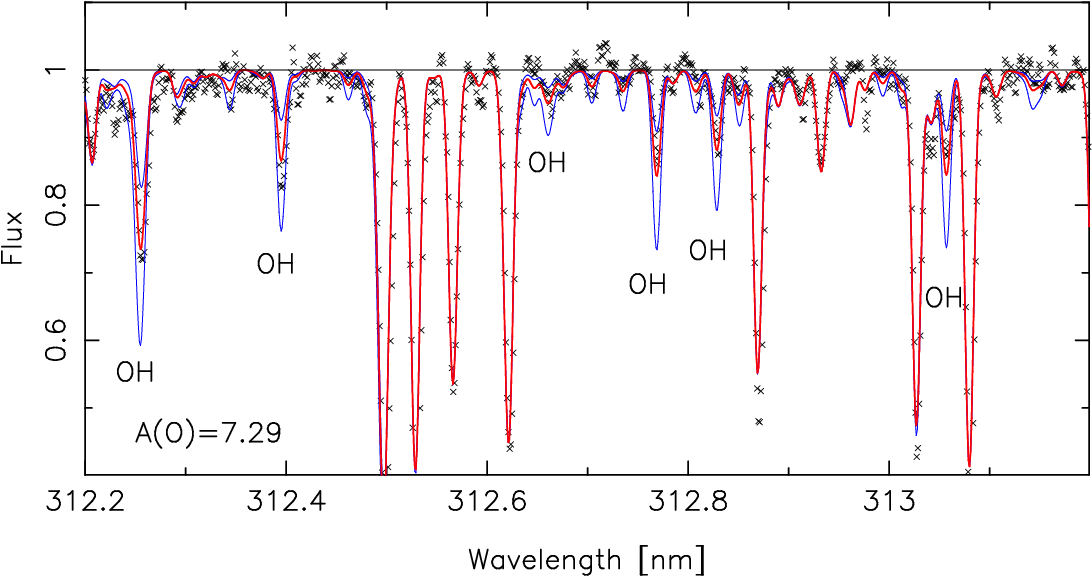}}
\caption[]{Observed profile of the CH and OH bands in \hdd. For the CH bands the 1D ETL synthetic profiles (blue thin lines) have been computed with A(C)= 0.0, 6.4, and 6.7. 
The absorption by the wing of H$\gamma$ has been included in the computation of the G-band.
The thick red line represents the best fit: A(C)=6.60 for the UV CH feature and A(C)=6.64 for the G-band. 
The synthetic profile of the OH band has been computed with A(O)= 7.0 and 7.6. The best fit is obtained with A(O)= 7.29
}
\label {CHOHlines}
\end{figure}

\subsection {Carbon and oxygen abundances} 

In \hd the C and O abundances can be derived from atomic lines and from the molecular CH and OH bands.

\subsubsection {C and O from atomic lines}
\begin{itemize}
\item {Carbon}

It is possible to determine the carbon abundance from a fit of the weak red \Cu\ lines near 910\,nm, the weak UV lines around 199 and 248\,nm and the very strong UV \Cu\ line at 193\,nm. In the three cases we estimate (Table  \ref{abund})  that the error in [C/H] is $\approx$ 0.1\,dex. 

 The red \Cu\ lines are located within a \Hdo\ telluric band. In this region we had several spectra observed at different dates (2002, 2003, 2004), and for each mean spectrum the relative position of the telluric lines was different. Thus we chose to measure each \Cu\ line from only those spectra in which telluric contamination was minimal. The 906.25 line was measured only on the mean 2003 spectrum, the 907.83 line could be measured on the 2002 and 2003 mean spectra, and the 911.18 line only on the 2004 spectrum. An example of the fit is given in Fig. \ref{C1red}. 
In Fig. \ref{C1strongUV} we present the fit of the wings of the strong \Cu\ line at 193.1\,nm.

\citet{FabbianAC06,FabbianNA09} computed  a non-LTE correction for \hdd\ of --0.2\,dex for the high excitation \Cu\ red lines. \citet{ZhaoMY16} and Mashonkina \pc\ 
computed a correction close to --0.1\,dex for the red \Cu\ lines and $\leq 0.01$ dex for the UV \Cu\ lines; we adopted these  corrections (Table \ref{abundlines}).
These new measurements point to a [C/Fe] ratio in \hd close to solar (Table \ref{abund}).

\item{Oxygen}

We derived the oxygen abundance in \hd from the oxygen triplet at 777\,nm  (Fig. \ref{Olines}).  The non-LTE correction ($-0.12$\,dex) was  adopted from the calculations of \citet{ZhaoMY16} and leads to  $\rm [O/H] = 7.16 \pm 0.1$. 
The resulting value  $\rm [O/Fe] = +0.65$ is in good agreement with the value obtained from the forbidden [\ion{O}{i}] line in the EMP giants by \citet{SpiteCP05},  and is in a reasonable agreement, inside the measurement errors, with the value obtained by \citet{ZhaoMY16} from the red \Ou\ lines (part of the difference indeed comes from the adopted [Fe/H] value).
\end{itemize}

\begin{figure}[h]
\resizebox{\hsize}{!}                   
{\includegraphics {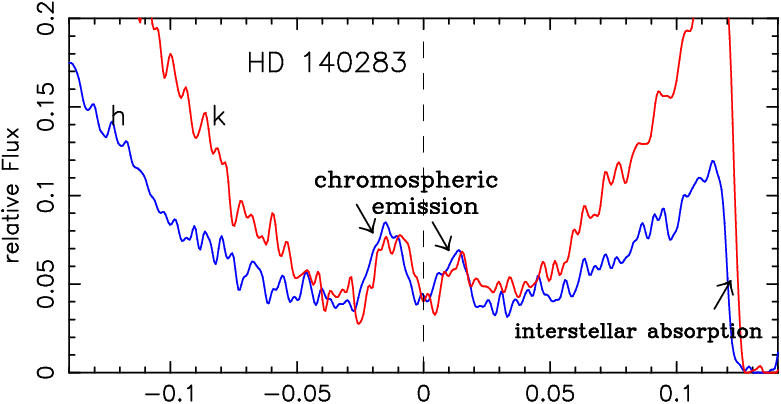}}
\resizebox{\hsize}{!}                   
{\includegraphics {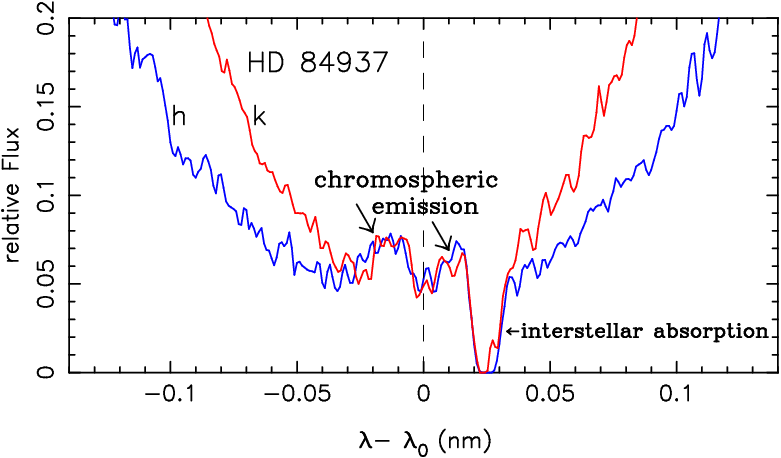}}
\caption[]{Centre of the the h and k \Mgd\ lines in the HST/STIS UV spectra of \hd and HD\,140283. The profiles of the h and k lines have been overplotted to show the emission at the centre of both lines indicating the existence of a chromosphere.
In \hd  an interstellar line, with a radial velocity close to the radial velocity of the star, perturbs the profile on the red side of the core. 
}
\label {mg2}
\end{figure}

\subsubsection {C and O from CH and OH molecular bands}

~~~~~~ --1D Computations  \\
-We computed the CH carbon abundance by fitting the profile of the CH feature at 314.3\,nm and of the G-band between 430.8 and 431.4\,nm.  Line lists for \CHdou ~and \CHtre~ \citep{MasseronPVE14} were included in the synthesis. 
The result is presented in Fig. \ref{CHOHlines}.  
For the UV feature the best fit corresponds to $\rm[C/H]=-1.90$\,dex, i.e., $\rm[C/Fe]=+0.35$\,dex, and for the G-band to $\rm[C/H]=-1.86$\,dex, i.e., $\rm[C/Fe]=+0.39$\,dex.  The CH band in \hd is very weak and thus the corresponding abundance is very sensitive to the position of the continuum and in this case the error on $\rm[C/H]$ is estimated to $\approx 0.15$\,dex.\\ 
-A fit of the OH band  between 312 and 314\,nm leads to [O/H]=--1.47 (Fig. \ref{CHOHlines}) and [O/Fe]=+0.78.

There is, in \hdd,  a rather good agreement between the abundances deduced from the atomic lines (Table \ref{abund}) and the molecular bands from 1D computations. However, in metal-poor stars the molecular lines are sensitive to  convective effects, neglected in the 1D computations, and we have to take them into account.

--3D Computations  \\
We computed 3D corrections to the 1D abundances determined for the G-band, 
and $14$ OH lines in the wavelength range $312.24-312.83$\,nm. We define a 3D correction, $\Delta_{\rm 3D}$, with the formalism given by \citet{GallagherCB16}, i.e., $\Delta_{\rm 3D}=A({\rm X})_{\rm 3D}-A({\rm X})_{\rm 1D}$, where the 1D synthesis is computed from LHD model atmospheres. 
We computed 3D profiles of CH and OH using the absolute carbon and oxygen abundances derived from the 1D non-LTE computations of the red \Cu\ and \Ou\ atomic lines and thus with a C/O ratio of $\approx 0.15$.
(Unlike the 1D computations, the 3D computations are sensitive to the C/O ratio) \footnote{${\rm X/Y}=N({\rm X})/N({\rm Y})=10^{\left[ A({\rm X})_*-A({\rm Y})_* \right]}$}.       

For the G-band we found a 3D correction of --0.28\,dex. The resulting C abundance, [C/H]= --2.14 or [C/Fe]= +0.11, is in very good agreement with the abundance deduced from the atomic lines (Table \ref{abund}). 
However, the OH lines were far stronger in 3D than in 1D. In fact,  owing to the low C/O ratio in the atmosphere of a typical metal-poor star like \hdd, the OH lines are mainly formed in the most external regions of the atmosphere, which are about 300\,K cooler in the 3D model than in the 1D model 
\citep[see Fig. 1 in][for a metallicity of --2]{GallagherCB16}.
In this case, the 3D correction reaches $-0.72$\,dex. Applying this 3D correction to the OH band would lead to a value of [O/H]=--2.19 ([O/Fe]=+0.06), a value hardly compatible with the observation of the atomic lines. For \hd the adopted 3D model seems to be too cool in the most outer layers. This could be due to the omission of a hot solar-like chromosphere in our 3D model.\\ 

From HST spectra of the \Mgd\ lines near 280\,nm, \citet{PetersonS97} 
showed that solar-like chromospheres routinely occur among turnoff metal-poor 
stars despite their relatively low degree of activity. From seven such 
stars of a wide metallicity range down to 1/300 solar whose radial 
velocities largely shifted the line cores away from the interstellar 
absorption, this work demonstrated that double-peaked emission cores 
were always present, and that the blue peak was usually stronger than the red 
as is also the usual case for the quiet Sun.
They concluded that ``while these data do not rule out magnetic fields they support an acoustic origin of chromospheric emission and show that relatively inactive solar-type stars of all ages have 
chromospheres whose characteristics are largely independent of metallicity''.
Subsequent work, based on observations of \Heu\ 1083\,nm 
absorption features, have confirmed these results \citep[e.g.,][]{TakedaT11}.
More work on this subject, based on the profiles of the \Mgd\ lines in HST/STIS spectra of several metal-poor stars, is ongoing (Peterson \& Tarbell in preparation). In Fig. \ref{mg2} we present the profile of the \Mgd\ h and k lines in our HST/STIS spectrum of \hd and in the
G0-14161 E230H spectrum of another classical extremely metal-poor star, HD\,140283 \citep{PetersonKA16}. In HD\,140283 the two peaks -- typical of a solar-like chromosphere -- are clearly visible; unfortunately, in \hd  an interstellar line with a radial velocity very close to the radial velocity of the star perturbs this profile on the red side, and the red peak of the emission is not clearly visible.

We also computed the 3D correction for two \Cu\ lines: the very strong line at 193.09\,nm 
(Fig. \ref{C1strongUV}) and the line at 199.36\,nm.  In both cases we found a correction lower than the measurement error (respectively +0.06 and -0.07).

\subsection {Abundances of classical $\alpha$ elements: magnesium and calcium}

For a more homogeneous comparison,  in the visible we derived the magnesium and calcium abundances from the equivalent widths of a set of lines already used in extremely metal-poor (EMP) main-sequence turnoff stars by \citet{BonifacioSC09}.  \\
~~ 

 For the optical \Mgu\ lines in Table \ref{abundlines}, we adopted the non-LTE correction computed by \citet{ZhaoMY16} and we found a mild overabundance of Mg (Table \ref{abund}: $\rm [Mg/Fe]_{LTE}= +0.30$ dex), as did \citet{GehrenSZ06} and \citet{ZhaoMY16}.
This value is also in good agreement with the value we deduce from the wings of the very strong \Mgd ~h and k lines around 280\,nm ($\rm [Mg/Fe]_{LTE}= +0.26$ dex). The non-LTE correction has not been computed for these UV lines. However, following \citet{AbiaMash04}, since \Mgd\ is the majority species,  departures from LTE are expected to be caused mainly by radiative b-b transitions, and no process seems to affect the \Mgd\ ground-state population. As a consequence, the non-LTE correction for the resonance h and k \Mgd\ lines should be negligible.  \\
~~
 Our LTE calcium abundances in Table \ref{abundlines} have been corrected for non-LTE effects following \citet{SpiteAS12}.
We have three lines in common with \citet{ZhaoMY16}, and the non-LTE corrections for these lines do not differ by more than 0.02\,dex. 
The Ca abundance found in \hd\ is in very good agreement with the mean abundances obtained, using the same set of lines,  for EMP turnoff stars with $\rm-3.6<[Fe/H]<-2.5$ \citep{BonifacioSC09,SpiteAS12}.

\subsection {Silicon abundance} 
The silicon abundance could be deduced from seven silicon lines in the UV and another one at 390\,nm. This last line was used to determine the Si abundance in the sample of EMP turnoff stars studied in \citet{BonifacioSC09}. \citet{ShiGM09} computed for this line a non-LTE correction of +0.05 dex. The non-LTE correction for the ultraviolet lines were computed by Mashonkina \pc\  and are less than 0.1\,dex (Table \ref{abundlines}).
Our derived overabundance of Si is then $\rm [Si/H]=-1.87 \pm 0.10$ or $\rm [Si/Fe]=+0.38$. 
It is interesting to note that this value is higher than the mean silicon overabundance derived in EMP dwarfs from the 390\,nm line alone: $\rm [Si/Fe]\approx +0.09 \pm 0.14$. But this mean value has not been corrected for non-LTE effects and, in these EMP turnoff stars with $\rm [Fe/H] \leq -3$, the non-LTE correction could reach 0.25 dex \citep{ShiGM09}. As a consequence the silicon abundance found in the UV in \hd is compatible with the results of \citet{BonifacioSC09}.
As for Mg and Ca, our [Si/H] value is  very close to the value found by \citet{ZhaoMY16}, and the small difference in [Si/Fe] is attributed to the adopted [Fe/H] values.

\subsection {Phosphorus abundance} 
\begin{figure}[h]
\resizebox{\hsize}{!}                   
{\includegraphics {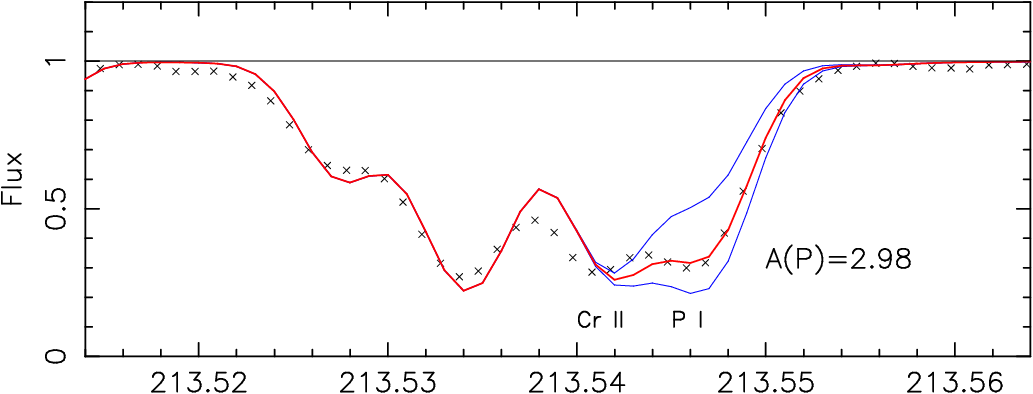}}
\resizebox{\hsize}{!}                   
{\includegraphics {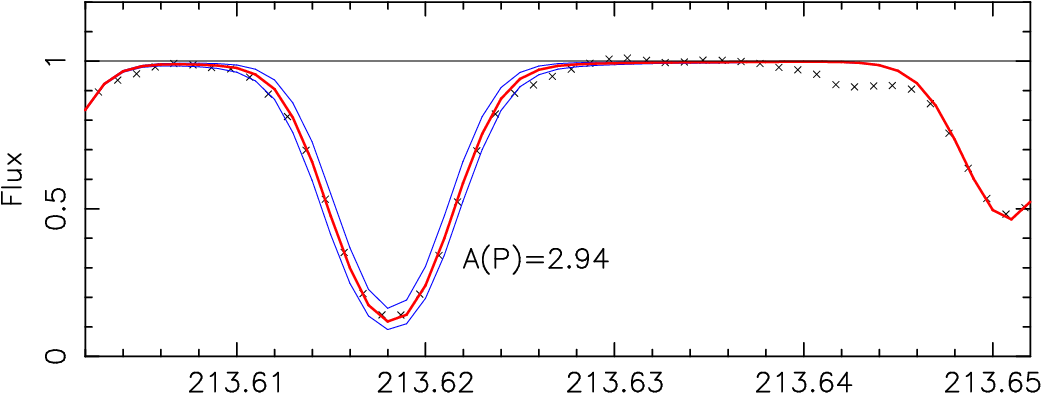}}
\resizebox{\hsize}{!}                   
{\includegraphics {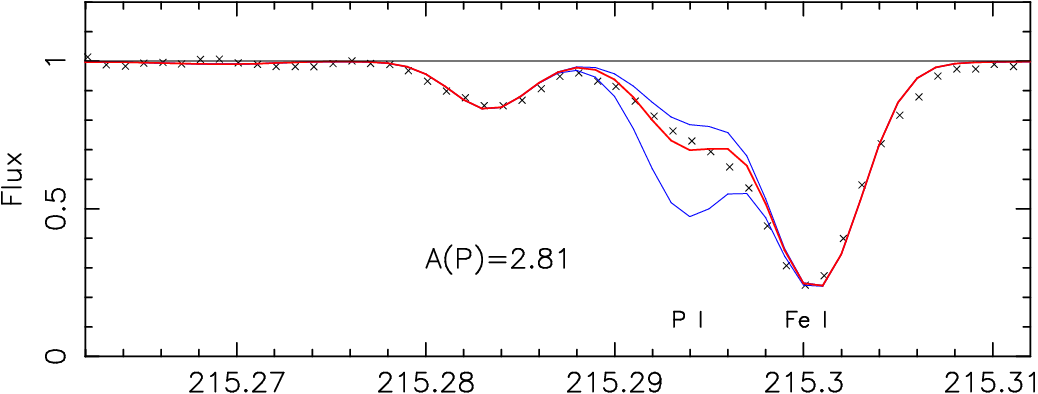}}
\resizebox{\hsize}{!}                   
{\includegraphics {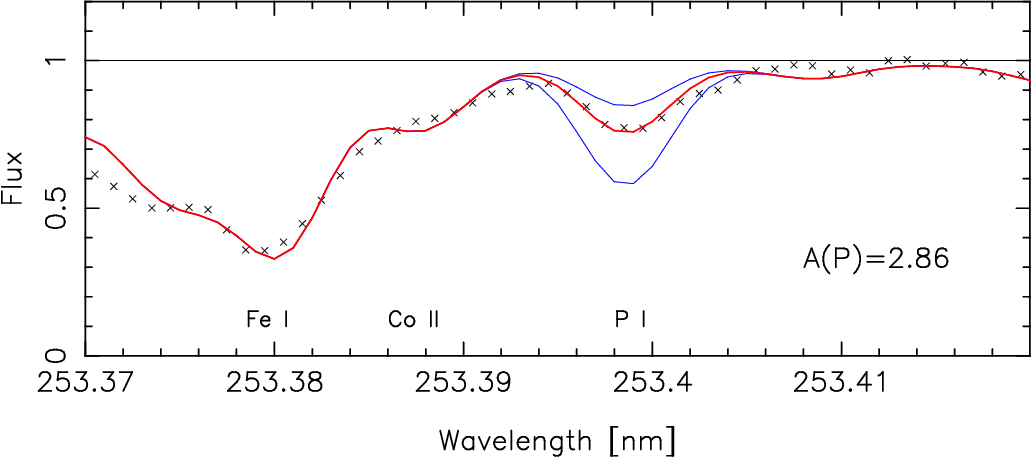}}
\caption[]{Comparison of the observed (crosses) and synthetic (curves) spectra in \hd in the region of the four P I lines analysed in this work. The red thick line marks the best-fit synthetic spectrum, obtained with the value of  A(P) indicated in the figure. The thin blue lines show synthetic spectra computed with A(P)=2.6 and A(P)=3.2.
As  is usually done, the \loggf\ values of the blending lines have been adjusted to give a more correct representation of the different features.}
\label {Plines}
\end{figure}

\begin{figure}[h]
\resizebox{\hsize}{!}                   
{\includegraphics {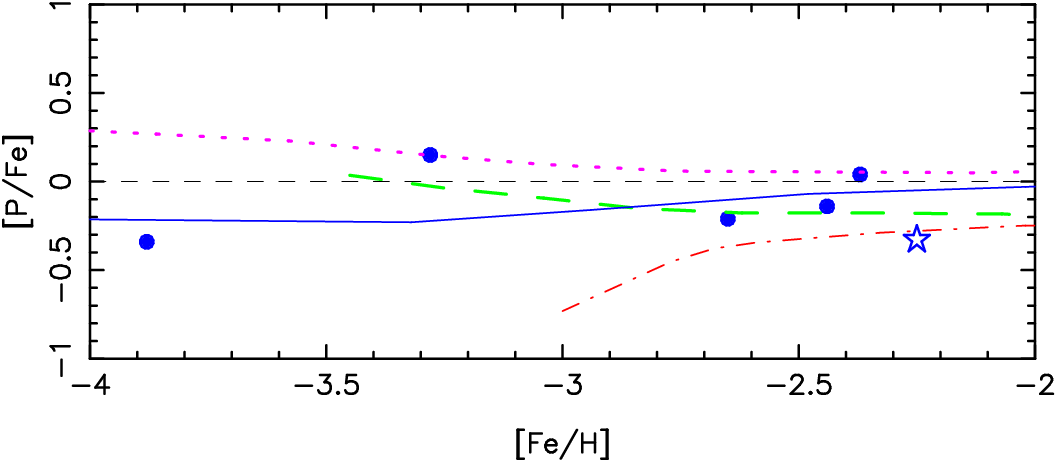}}
\caption[]{[P/Fe] vs. [Fe/H] is shown for the five stars analysed by  \citet{RoedererJT14} and \citet{RoedererPB16} with $\rm\ [Fe/H] < -2.0$ (blue dots), plus our result for \hd (blue star symbol). The theoretical predictions of \citet{TimmesWW95} (red dot-dashed line), \citet{GoswamiP00} (pink dotted  line), \citet{KobayashiKU11} (blue thick line), and \citet{CescuttiMC12} (green dashed line)  are also represented.}
\label {p-mpoor}
\end{figure}

The phosphorus abundance is derived from five ultraviolet lines of the multiplets 4(uv), 8(uv), and 9(uv), leading to $\rm [P/Fe] = -0.32$. The statistical error reaches 0.07\,dex. Some other P lines are visible in the spectrum, but these lines are so severely blended that they cannot be used to determine the P abundance with precision.
In Fig. \ref{Plines} we show the fit of the observed lines with the computed spectra. 
The  value of [P/Fe] in \hd confirms the low value of [P/Fe] in metal-poor stars (Fig. \ref{p-mpoor}).   \\ 
Recently R. Kurucz has published new \loggf ~values for the P lines in the UV and in the red 
\footnote{ {\tt http://kurucz.harvard.edu/linelists.html} see file {\tt GFALL}.}. 
These new values are about 0.1\,dex higher than the old ones. Consequently, their adoption here would reduce the  [P/Fe] values of Tables \ref{abundlines} and \ref{abund} by about 0.1\,dex.

To date, no non-LTE calculations have been performed for \Pu~ features; following \citet{JacobsonTF14} they are expected to be small in this type of star. 

\begin{figure}[h]
\resizebox{\hsize}{!}                   
{\includegraphics {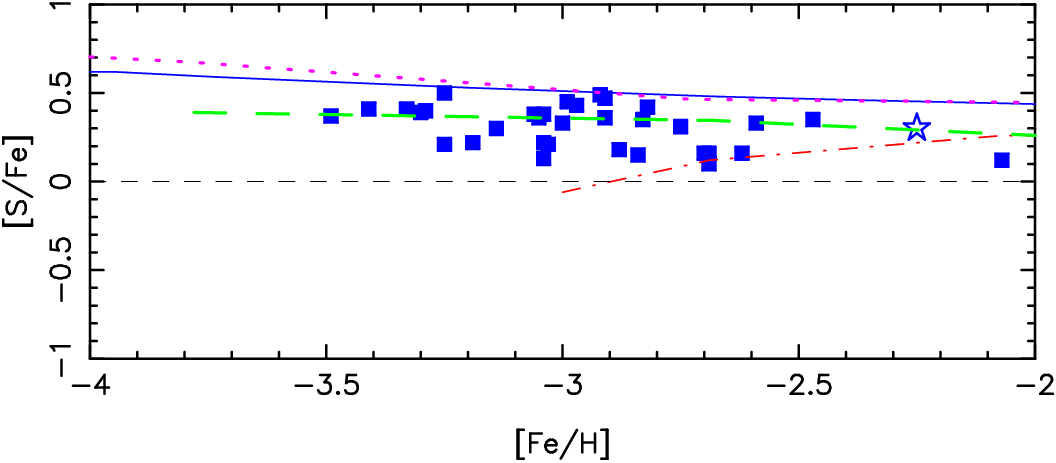}}
\caption[]{[S/Fe] vs. [Fe/H] for metal-poor stars with $\rm\ [Fe/H] < -2.0$ (blue squares), plus our result for \hd (blue star symbol). The theoretical predictions of \citet{TimmesWW95} (red dot-dashed line), \citet{GoswamiP00} (pink dotted  line), \citet{KobayashiKU11} (blue thick line),  and \citet{Matteucci16} (green dashed line)  are also represented.}
\label {s-mpoor}
\end{figure}

\begin{figure}[h]
\resizebox{\hsize}{!}                   
{\includegraphics {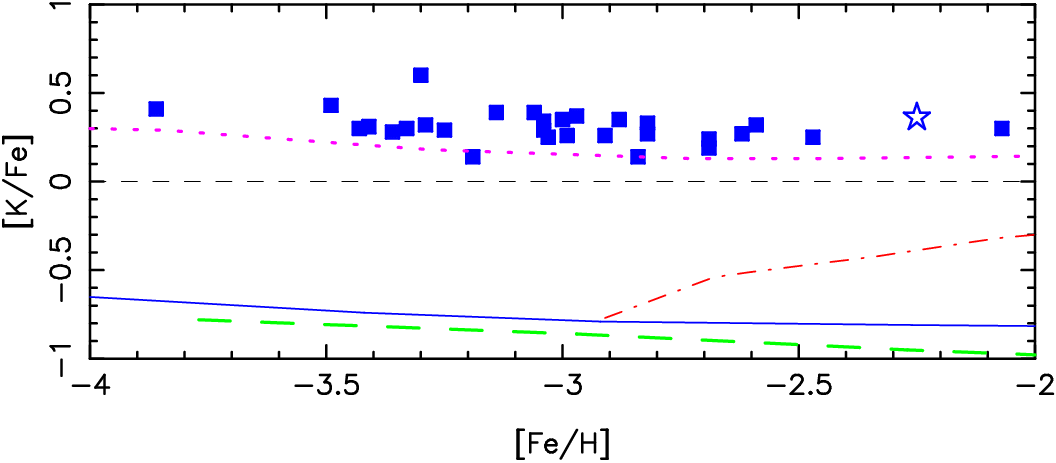}}
\caption[]{[K/Fe] vs. [Fe/H] for metal-poor stars with $\rm\ [Fe/H] < -2.0$ (blue squares), plus our result for \hd (blue star symbol). The theoretical predictions of \citet{TimmesWW95} (red dot-dashed line), \citet{GoswamiP00} (pink dotted  line), \citet{KobayashiKU11} (blue thick line),  and \citet{Matteucci16} (green dashed line) are also represented.}
\label {k-mpoor}
\end{figure}

\subsection {Sulphur abundance} 
In \hdd, the sulphur abundance could be derived from the profile of two UV features at 190.03 and 191.47\,nm and from the profiles of the red multiplet 1 (Table \ref{abund}). 
This multiplet is contaminated by a telluric band of \Hdo. We measured the S lines only on spectra where they were not blended by a telluric line. The 921.2 line could be measured on the 2002, 2003, and 2004 spectra, but the  other two lines (922.8 and 923.7m) only on the 2002 and 2003 spectra. 
We obtained good agreement between the abundances deduced from the two different sets of lines: [S/Fe]=+0.42 dex from the UV lines and [S/Fe]=+0.50 from the red lines.\\
It is well known that non-LTE effects influence 
the profiles of the red multiplet 1. Following \citet{TakedaHT05}, the non-LTE correction is about --0.2 dex for the model parameters of \hdd. 
However,  following \citet{SpiteCA11}, based on a new model atom of sulphur by  \citet{Korot08,Korot09a,Korot09b} and new photoionisation rates, the NLTE correction for the model of \hd reaches --0.4\,dex. On the other hand, it has been computed that in the case of this multiplet the 3D correction is not negligible and reaches +0.1\,dex \citep{SpiteCA11}. This   small but positive correction has been added to the NLTE correction. Finally, we find from the red multiplet 1, [S/Fe]=+0.20\,dex. 

As far as we are aware, non-LTE corrections of the UV features have never been computed, but they are probably as small as for C and Si.   
We estimate a sulphur overabundance in \hd of $\rm [S/Fe] \approx 0.3\,dex$, in good agreement with the mean sulphur overabundance found in mid metal-poor stars \citep{Francois86} and in EMP stars \citep{SpiteCA11}.

\subsection {Potassium abundance} 
The abundance of K has been deduced from the red lines at 766.4899\,nm and 769.8964\,nm. Following \citet{TakedaZC02}, the non-LTE correction for the model parameters of \hd reaches --0.3\,dex, but \citet{ZhaoMY16} has computed a correction of  --0.2\,dex and we adopted this latter value. 
As a consequence we find for this star a ratio $\rm [K/Fe]=0.37 \pm 0.10$, in excellent agreement with the value found for the EMP stars  by \citet{TakedaKM09}  and \cite{AndrievskySK10}.

\section {Discussion and conclusion}
 
Our analysis of high quality spectra for \hdd, notably the HST ultraviolet spectra as well as ground-based optical/near-infrared spectra, has allowed us to refine the abundances of several light elements between C and Ca in this star, and provide the first measurements of its P, S, and K abundances and a more firmly established Si abundance. Accurate abundances of such key elements are vital in reference stars like \hdd\ in order  to better interpret chemical evolution models and supernovae enrichment in the Galactic halo.

The presence of emission-core reversal at the centre of the very strong \Mgd\ resonance lines confirms the existence of a chromosphere around \hd \citep{PetersonS97}.

The abundances of C and O deduced from the atomic lines are in good agreement with the abundances obtained in the EMP stars  \citep{BonifacioSC09,SpiteCP05}.\\ 
The C abundance deduced from the CH molecular band after the 3D-1D correction is in very good agreement with the abundance deduced from the atomic lines (Table \ref{abund}).\\
The oxygen abundance deduced from the OH molecular band computed with 1D models agree with the abundances deduced from the atomic lines; however, when 3D computations are done, the abundance of oxygen deduced from the OH band is incompatible with the abundance deduced from the atomic lines.
In 3D computations, at variance with 1D computations,  the OH band is formed predominantly in shallow atmospheric layers, and since 3D models are cooler at the surface than 1D models, OH lines 
are highly strengthened. However,  current 3D models for metal-poor turnoff stars do not include the rise in temperature exhibited by a solar-like chromosphere. 
These models have been built only for the Sun, but similar models would be useful in order to analyse molecular features in classical metal-poor turnoff stars.

The classical $\alpha$ elements Mg, Si, and Ca  have about the same overabundance in \hd as observed in the EMP turnoff stars studied in \citet{BonifacioSC09}.
\hd is also mildly enhanced in K, as are the other metal-poor stars.
 
--The ratio [P/Fe] is found to be slightly below the solar value, as is typically seen in very metal-poor stars (Fig.\ \ref{p-mpoor}) and more generally for the light odd-Z elements like Na and Al. \citet{MolaroLD01} also measured  [P/Fe] in a dust free Ly-$\alpha$ system that has the same metallicity as \hdd, and they found [P/Fe]\,=\,--0.27, about the same ratio as observed in \hd (Table \ref{abund}).  \\
In Fig.\ \ref{p-mpoor} we show the P measurements of \citet{RoedererJT14,RoedererPB16}. The two most metal-poor stars in the figure are \bd~ ([Fe/H]=$-3.88$) and G\,64-12, ([Fe/H]=$-3.28$). They are both carbon-enhanced metal-poor stars  (CEMP) following \citet{RoedererPB16} and \citet{PlaccoBR16}, but in these stars the neutron-capture elements are not enhanced. They are ``CEMP-no'' stars and thus the composition of their atmosphere is supposed to reflect the abundance of the cloud that formed the star. This cloud would have been enriched by a faint supernova (SN) providing carbon and the lighter elements,  but the elements heavier than Mg are not affected  \citep[see, e.g.,][]{BonifacioCS15}.
As a consequence, the P abundance in these stars should be comparable to the P abundance in classical metal-poor stars (i.e., metal-poor stars without C enhancement) and we include them in the figure.  \\
~~
According to \citet{WoosleyW95} and \citet{WestH13}, the isotope $\rm^{31}P$ is produced mostly in SN\,II in both O and Ne burning shells, and its ejection is not very        sensitive to the explosion mechanism. 
Taking into account core-collapse supernovae and hypernovae, \citet{KobayashiUN06,KobayashiKU11} produced models of chemical evolution for the halo. We overplotted their best model to the very few available P data points (Fig.\ \ref{p-mpoor}).
Also plotted in this figure are the metallicity-dependent model of \citet{TimmesWW95}, the model of \citet{GoswamiP00} using yields from \citet{WoosleyW95} at variable metallicities,  and   model 4 of \citet{CescuttiMC12}.
The best fit seems to be obtained with the   \citet{KobayashiKU11} model.
However in the interpretation of the abundance trends we have to be very prudent. We note that NLTE computations have not been done for this element and that following \citet{CaffauAK16}  for the disk stars there is a shift of 0.23\,dex between the UV-HST and the near-infrared measurements of the P abundance (see their Fig.~5). This disagreement could be the result of different non-LTE corrections for the UV and near-infrared lines but could be also the result of systematic errors in the $gf$ values.

--Sulphur is an $\alpha$ element mainly produced in massive supernovae or hypernovae. In Fig. \ref{s-mpoor} we compare the observed [S/Fe] ratios to the abundance ratios predicted by the models of \citet{GoswamiP00}, \citet{KobayashiKU11} and \citet{Matteucci16}. 
The best fit is obtained with the model of \citet{Matteucci16}.

--In the very metal-poor stars, potassium (an odd element: Z = 19) is enhanced relative to iron (Fig. \ref{k-mpoor}) as are the (even) $\alpha$ elements. 
The models  struggle to represent the abundance of K in the Galactic halo. The best fit (Fig. \ref{k-mpoor}) is achieved by the model of \citet{GoswamiP00}; however, these authors point out that this good representation could be fortuitous and might be due to a factor of two reduction of the Fe yields  combined with the adopted IMF \citep{KroupaTG93}.
Following \citet{KobayashiKU11} the underproduction of K in the theoretical massive supernovae/hypernovae could be the consequence of a neglect of the $\nu$ process. When the core of massive stars collapses, a large flux of neutrinos is emitted and these neutrinos interact with heavy elements through neutral-current reactions producing in particular F and K.

The HST has permitted an extension of the pattern of the light elements produced by the explosion of the progenitor of a very old, metal-poor, unevolved star whose composition is highly representative of the earliest stages of prior massive star nucleosynthesis. These elements are a key to a better understanding of the nucleosynthetic processes that have formed the elements in the early massive supernovae/hypernovae. It is very important that  future  access to the UV region of the spectra be preserved. This region is a very powerful  tool for the study of the early phases of the Galactic evolution.

\begin {acknowledgements} 
We are grateful to Lyudmila Mashonkina for a very  useful and competent report and for the communication of several unpublished non-LTE corrections. We also thank  Piercarlo Bonifacio who computed the ATLAS9 model and WIDTH profiles for us to check the \Cu\ lines.
This work was supported by the ``Programme National de Physique Stellaire'' and the ``Programme National de Cosmologie et Galaxies'' (CNRS-INSU). 
Partial support for R. Peterson was provided by NASA under HST GO-14161. R. Peterson thanks her GO-14161 coinvestigators Tom Ayres for his assistance in acquiring and reducing the HST spectra and Robert Kurucz for his advice in analysing them.
A. J. Gallagher acknowledges the funding by FONDATION MERAC and the matching fund granted by the Scientific Council of Observatoire de Paris. B. Barbuy was partially supported by FAPESP, CNPq, and CAPES.
\end {acknowledgements}

\bibliographystyle{aa}
{}

\end{document}